\documentclass[a4paper,fleqn,usenatbib]{mnras}

\usepackage{newtxtext,newtxmath}
\usepackage[normalem]{ulem}

\usepackage[T1]{fontenc}
\usepackage{ae,aecompl}

\usepackage{graphicx}	
\usepackage{amsmath}	
\usepackage{amssymb}	
\usepackage{natbib}	

\title[Praesepe white dwarfs]{Praesepe white dwarfs in \textit{Gaia}\,DR2}

\author[Salaris \& Bedin]{
M. Salaris$^{1}$\thanks{E-mail: M.Salaris@ljmu.ac.uk}
and L.~R. Bedin$^{2}$
\\
$^{1}$Astrophysics Research Institute, Liverpool John Moores
University,146 Brownlow Hill, Liverpool L3 5RF, UK\\ $^{2}$Istituto
Nazionale di Astrofisica -- Osservatorio Astronomico di Padova, Vicolo
dell'Osservatorio 5, I-35122 Padova, Italy\\ }

\date{Accepted 2018 November 30. Received 2018 November 14; in original form 2018 October 04}
\pubyear{2015}

\begin{document}
\label{firstpage}
\pagerange{\pageref{firstpage}--\pageref{lastpage}}
\maketitle

\begin{abstract}
  We have exploited \textit{Gaia}\, Data Release~2 to study white
  dwarf members of the Praesepe star cluster. We recovered eleven
  known white dwarf members (all DA spectral type) plus a new cluster
  WD never identified before.  Two of the eleven known DA objects did
  not satisfy all quality indicators available in the data
  release. The remaining nine objects of known spectral type have then
  been employed to determine their masses (average error of 3--5\%)
  and cooling times (average uncertainty of 5--7\%), by fitting
  cooling tracks to their colour-magnitude diagram. Assuming the
  recent \textit{Gaia}\, Data Release~2 reddening and main sequence
  turn off age estimates derived from isochrone fitting, we have
  derived progenitor masses and established the cluster initial-final
  mass relation.  We found consistency with the initial-final mass
  relation we established for eight Hyades white dwarfs, also
  employing \textit{Gaia} data. We have investigated also the effect
  on the derived initial masses of using self-consistently different
  sets of stellar models and isochrones for determining cluster age
  and white dwarf progenitor lifetimes.
 According to our established Hyades+Praesepe initial-final mass
 relation, recent sets of stellar evolution calculation that model the
 full asymptotic giant branch phase do on average underpredict the
 final white dwarf masses, in the initial mass range covered by the
 Praesepe and Hyades observed cooling sequence.
These results depend crucially on the assumed reddening for the
   cluster. To this purpose, we have also discussed the case of
   considering the traditional zero reddening for Praesepe, instead of
   $E(B-V)$=0.027 derived from isochrone fitting to the \textit{Gaia}
   colour magnitude diagram.
\end{abstract}

\begin{keywords}
open clusters and associations: individual
(Praesepe) -- stars: evolution -- stars: mass loss -- white dwarfs
\end{keywords}



\section{Introduction}

Theoretical calculations of the relationship between the initial (main
sequence) mass and the final carbon-oxygen (CO) white dwarf (WD) mass
for low- and intermediate mass stars is still challenging.  This stems
from the poorly modelled efficiency of mass-loss in stellar model
calculations, and uncertainties in the evolution of the mass size of
CO cores during the asymptotic giant branch (AGB) phase \citep[see,
  e.g.][]{iberen83, domi96, kl14}.

This state of affairs is problematic, because the initial-final mass
relation (IFMR) is an essential input for several astrophysical
problems. Obviously, location and shape of cooling sequences in
colour-magnitude diagrams (CMDs) and the associated WD luminosity
functions -- sometimes employed to age date stellar populations-- are
affected by the IFMR, but also chemical evolution histories of stellar
populations, their mass-to-light ratios, modelling of stellar feedback
in galaxy formation simulations \citep[e.g.][]{ak15}, Type Ia
supernova rate estimates \citep[e.g.][]{greggio} do depend on the
choice of the IFMR.

To overcome these shortcomings of stellar evolution models,
semi-empirical methods have been devised to establish the IFMR
\citep[see, e.g.,][for recent examples]{weide00, fer05, catalan08,
  kalirai09, ssw09, williams09, cumm15, cummings18}.

Semi-empirical IFMR techniques are still largely based on WDs hosted
by star clusters. Theoretical analyses of WD spectra provide surface
gravity $g$ and effective temperature $T_{\rm eff}$, by simultaneous
fitting of Balmer line profiles of DA WDs, employing high-resolution
observed and synthetic spectra.  For a given $g-T_{\rm eff}$ pair,
grids of theoretical WD models then provide the WD mass ($M_{\rm f}$)
and cooling age ($t_{\rm cool}$).

At the same time, theoretical isochrone fits to the main sequence (MS)
turn-off luminosity of the cluster CMD give the cluster age ($t_{\rm
  cl}$). The difference $t_{\rm cl} - t_{\rm cool}$ corresponds to the
lifetime of the WD progenitor until the start of the WD cooling
($t_{\rm prog}$). Finally, mass-lifetime relationships from
theoretical stellar evolution tracks provide an initial progenitor
mass ($M_{\rm i}$) from $t_{\rm prog}$ (the uncertain AGB and post-AGB
lifetimes can be neglected, because their duration is negligible
compared to the duration of the previous phases).

The high precision astrometry and three-band photometry (${\rm G,
  G_{BP}, G_{RP}}$) of \textit{Gaia} Data Release 2 (DR2) has enabled
to build CMDs of the closest open clusters \citep{gaiaclust}, that
display exquisitely defined sequences.  The distance modulus corrected
CMD of the Hyades cluster, for example, has typical errors (including
the parallax error contribution) of a few mmag in all three filters,
including the WD cooling sequence \citep[see, e.g.,][hereafter
  Paper~I]{sb18}.

Taking advantage of the DR2 parallaxes and photometry, in Paper~I we
have determined precise masses and cooling times (typically 1-3\%
precision) of the eight confirmed DA WDs Hyades members, by fitting
theoretical cooling tracks to the observed CMD. When absolute
magnitudes and colours are accurately known, this technique works
well, and is somewhat complementary to the spectroscopic one. In case
of CMD fitting, theoretical WD cooling sequences are used in
conjuction with --this time-- low-resolution synthetic spectra, needed
to calculate the appropriate bolometric corrections and colours.

The Hyades IFMR was then established from the knowledge of the cluster
age determined from the MS turn-off by \citet{gaiaclust}, employing
the DR2 CMD. WDs in the Hyades cover a range of masses corresponding
to $M_{\rm i}$ between $\sim 2.5$ and $\sim 4.0 M_{\odot}$, an {\sl
  interesting} mass range from the point of view of stellar
evolution. The lower limit corresponds approximately to stars just
beyond the threshold for He-ignition in a non-degenerate core. This
means that above $\sim 2.5 M_{\odot}$ the He-core mass at the start of
core He-burning starts to increase with increasing $M_{\rm i}$ (at
lower masses the He-core mass at He-ignition is approximately constant
with $M_{\rm i}$ because of the electron degeneracy).  The upper limit
of this mass range corresponds approximately to the onset of the
second dredge-up during the early-AGB phase. The second dredge-up
moderates the increase of $M_{\rm f}$ with increasing $M_{\rm i}$. As
a result, theoretical IFMRs predict for this initial mass range a
steeper slope in the $M_{\rm i}$-$M_{\rm f}$ diagram, compared to
lower and higher mass ranges \citep[see, e.g., predictions from models
  by][]{mist, parsecmimf}.

In this paper we consider \textit{Gaia} DR2 data for the Praesepe
cluster --roughly coeval and with the same metallicity of the Hyades--
that include a well defined WD sequence.  We apply the same techniques
employed in Paper~I to determine the cluster IFMR. Putting together
the Hyades and Praesepe IFMR derived from the DR2 data, allows us to
test theoretical IFMRs in this important $M_{\rm i}$ range.  In
Paper~I we assessed the --negligible-- effect on the WD masses and
cooling times (and the derived IFMR) of employing three independent
sets of WD cooling models. Here we explore the effect of employing
independent sets of stellar evolution models/isochrones to determine
progenitor masses, and their impact on the IFMR.

The plan of the paper is as follows. Section~\ref{data} describes
briefly the Praesepe WD sample, whilst section~\ref{analysis}
describes our derivation of the WD masses, cooling times, and
comparisons with previous independent determinations.
Section~\ref{ifmr} presents our derivation of the IFMR with associated
errors employing the cluster age and reddening determined from
isochrone fitting to the \textit{Gaia} main sequence and turn off CMD,
the effect of using different sets of stellar models/isochrones to
determine $M_{\rm i}$, and comparisons with theoretical IFMR
predictions. In addition, we also rederive the IFMR using this time
the recent cluster age determined from Johnson photometry by
\citet{cummings18}, considering the traditional zero reddening for
this cluster.  A summary and discussion follow in
Sect.~\ref{conclusions}.

\section{Data}
\label{data}

The 932 Praesepe members considered for this work are those defined
and released by \citet{gaiaclust}. We translated apparent G-band
magnitudes and observed $\rm {(G_{BP}-G_{RP})}$ colours to absolute
magnitude and reddening-corrected colours employing the individual DR2
parallaxes, $E(B-V)$=0.027 as derived (together with the age)
  from isochrone fitting by \citet{gaiaclust}.  We notice that this
  value agrees with the reddening determined by \citet{taylor}; the
  associated formal error on \citet{taylor} determination is
  negligible, equal to 0.004~mag (see Sect.~\ref{c18} for more on
  Praesepe reddening).

We have added to the published DR2 parallaxes a zero-point correction
by 0.03~mas, following \citet{lindegr18}. The effect of this
correction on the absolute magnitudes is however almost negligible.
Extinction coefficients for the three \textit{Gaia} photometric
filters have been derived as described in Sect.~2.2 of
\citet{gaiaclust} -- see their Eq.~1.

\begin{figure}
	\includegraphics[width=\columnwidth]{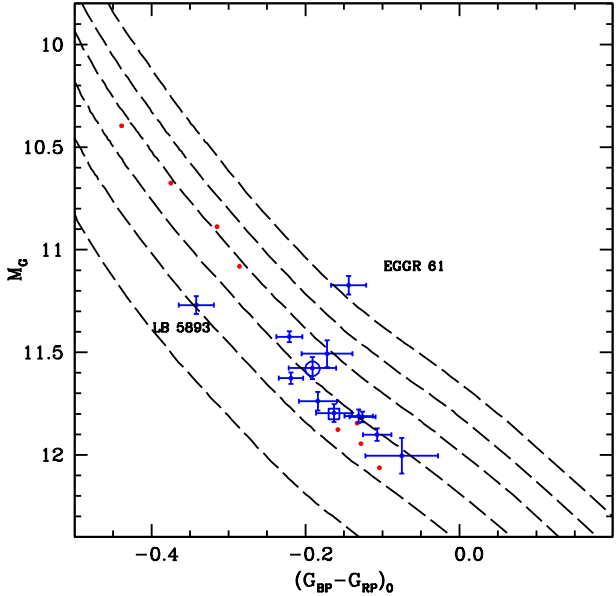}
        \caption{\textit{Gaia} DR2 CMD --distance modulus and
          reddening corrected-- of the sample of 12 Praesepe WD
          members (blue filled circles with error bars) together with
          the \citet{bastiwd} H-atmosphere cooling tracks for masses
          equal to 0.54, 0.61, 0.68, 0.77, 0.87 and 1.00~$M_\odot$
          (see text for details).  Error bars include the DR2 quoted
          photometric errors and the contribution from the parallax
          error.  The object enclosed within an open circle is a
          magnetic WD (see text). The object enclosed within an open
          square is a new member WD (see text for details).  Filled
          red circles without error bars denote the Hyades WDs of
          Paper~I.  The two labelled Praesepe WDs are {\sl peculiar}
          objects discussed in the text.}
    \label{fig:sample}
\end{figure}

The resulting cooling sequence in the \textit{Gaia} CMD is displayed
in Fig.~\ref{fig:sample}. The 1~$\sigma$ error bars take into account
both photometric and, for the absolute G-band magnitudes, parallax
errors.  Parallax fractional errors are typically around 4\%, while
1$\sigma$ errors on ${\rm M_G}$ and colours are in the range 0.03-0.07
and 0.02-0.03~mag, respectively.

The sequence is populated by twelve WDs, eleven of which are already
known cluster members \citep[see][]{at, clav01, dob04, dobbie06,
    casewell}.  These known members are all DA objects, and are
  all listed in the Montreal White Dwarf Database \citep{mwdd}.
\citet{casewell} argued that one of these eleven objects (WD~0837+218)
is possibly a non-member, but it is found to be a cluster member by
\citet{gaiaclust} in their analysis of star clusters in
\textit{Gaia}~DR2.  The new WD of unknown spectral type has the DR2
identifier \#662998983199228032.

Figure~\ref{fig:sample} shows also the Hyades cooling sequence of
Paper~I (red dots), as well as \citet{bastiwd} theoretical DA WD
cooling tracks, for masses $M_{\rm WD}$ between 0.54 and
1.0~$M_\odot$.  The cooling tracks are calculated for CO cores
\citep[see][for details about the CO stratification]{bastiwd} and
thick H layers (mass thickness equal to $10^{-4}M_{\rm WD}$, on top of
a $10^{-2}M_{\rm WD}$ He layer). The theoretical absolute
  magnitudes in the \textit{Gaia}\,DR2 filter system were determined
  from the model bolometric luminosities by applying bolometric
  corrections kindly provided by P. Bergeron \citep[private
    communication, see][]{hb06, tbg11}.

As expected, due to the cluster similar ages and metallicities
\citep[e.g.][]{gaiaclust}, Hyades and Praesepe WDs appear to follow a
single cooling sequence, with the exception of the two labelled
Praesepe objects. They are too blue (mass too high) and too red (mass
too low) compared to the combined cooling sequence of the two
clusters.

To investigate whether these peculiar colours are due to issues with
DR2 data, we have checked the quality indicators available in
\textit{Gaia}\,DR2\footnote{\url{https://gea.esac.esa.int/archive/documentation/GDR2/}}
for the whole sample of Praesepe WDs. As done in Paper~I for the
Hyades WDs, we have considered \texttt{visibility\_periods\_used,
  astrometric\_matched\_observations, astrometric\_gof\_al,
  astrometric\_excess\_noise, astrometric\_n\_good\_obs\_al,
  astrometric\_n\_bad\_obs\_al}, plus all the estimated errors on
motions, positions and magnitudes for all filters, and compared with
their average values for magnitude intervals.  Finally, we also
inspected the tests for well measured objects defined in Eqs. (C.1)
and (C.2) of \citet{lindegr18}.

All these indicators --for the whole WD sample-- appear to be
reasonably well measured, with the only exceptions of EGGR~60
(\#661311267210542080) --the faintest object in
Fig.~\ref{fig:sample}-- and EGGR~61 (\#661297901272035456). These
objects did not pass the test defined by Equation (C.2) of
\citet{lindegr18}, that flags problem with the ${\rm G_{BP}}$ and
${\rm G_{RP}}$ photometry.
For this reason we won't consider these WDs in the analysis that follows.

Also, According to \citet{casewell}, the WD LB~5959
(\#659494049367276544)--that sits within the well defined cooling
sequence, at an absolute G magnitude equal to 11.74 and ${\rm
  (G_{BP}-G_{RP})_{0}}$=$-$0.18-- is a radial velocity variable.
However, they could not find compelling direct evidence of any cool
companion from other observational datasets.  Follow-up
  observations presented by \citet{casewell12} confirmed the radial
  velocity variability, leading to the conclusion that the probable
  companion is a 25-30~$M_{\rm Jup}$ T~dwarf.  We find that all
\textit{Gaia}\,DR2 parameters for this objects --and in particular the
estimated errors in magnitudes-- do not show any indication of this
being a binary system.

We have also highlighted with an open circle the object EGGR~59, that
is a known magnetic WD with field strength of approximately 3~MG
\citep{casewell, ferr15}.  This WD lies well within the combined
Hyades-Praesepe cooling sequence, and we will retain it in our
analysis.  We notice that omitting this object from the analysis that
follows would not alter the main results of this paper.

Table~\ref{tab:dataWD} summarizes parallaxes (with the zero-point
correction applied), their fractional errors, absolute magnitudes in
the G filter (${\rm M_G}$), as well as the dereddened $\rm
{(G_{BP}-G_{RP})_{0}}$ colours and associated $1\sigma$ errors (taking
into account also the errors on the parallax) for the nine known
Praesepe WDs considered in our analysis.

\section{Analysis}
\label{analysis}

Masses and cooling times of the Praesepe WDs listed in
Table~\ref{tab:dataWD} have been determined as described in Paper~I
for the Hyades sample.  Interpolations amongst the \citet{bastiwd}
cooling tracks shown in Fig.~\ref{fig:sample} to match ${\rm M_G}$ and
${\rm (G_{BP}-G_{RP})_{0}}$ of each individual WD, provide mass and
cooling age ($M_{\rm f}$ and $t_{\rm cool}$), reported in
Table~\ref{tab:dataWD}. The associated errors have been estimated by
generating for each object one thousand synthetic ${\rm M_G}$ and
${\rm (G_{BP}-G_{RP})_{0}}$ pairs, with Gaussian distributions
(assumed to be independent) centred around the measured values, and
1$\sigma$ widths equal to the errors on these quantities reported in
Table~\ref{tab:dataWD}.  Mass and cooling times for each synthetic
sample were determined from the WD tracks, and the 68\% confidence
limits calculated.

\begin{table*}
	\centering
	\caption{Data about the 9 known DA Praesepe WDs shown in
          Fig.~\ref{fig:sample}, after discarding EGGR~61 and EGGR~60
          (see text for details). We display, from left to right, WD
          name, \texttt{Identifier:\,Gaia\,DR2}, parallax (in mas),
          parallax fractional error, absolute ${\rm G}$ magnitude with
          error (including the contribution from the parallax error),
          colour with associated error, logarithm of the cooling time
          (in years) and error, mass (in solar units) and associated
          error.}
	\label{tab:dataWD}
	\begin{tabular}{rrccccccc} 
		\hline
		Name & \texttt{Identifier:\,Gaia\,DR2}&$\pi$ &$\sigma_{\pi}/\pi$ &${\rm M_G}\pm \sigma$ & ${\rm (G_{BP}-G_{RP})_{0}\pm\sigma}$& ${\rm log}(t_{\rm cool})\pm\sigma$&$M_{\rm f}\pm\sigma$\\
               (1) & (2) & (3) & (4) & (5) & (6) & (7) & (8) \\
		\hline
              LB~5893     & \texttt{661270898815358720} & 5.447  &0.040  &11.27$\pm$0.04  & $-$0.34$\pm$0.02 &  8.10$\pm$0.03 &  0.87$\pm$0.04 \\
              EGGR~59     & \texttt{664325543977630464} & 5.422  &0.040  &11.58$\pm$0.05  & $-$0.19$\pm$0.03 &  8.40$\pm$0.04 &  0.75$\pm$0.04 \\
              LB~1876     & \texttt{661353224747229184} & 5.850  &0.035  &11.63$\pm$0.03  & $-$0.22$\pm$0.02 &  8.40$\pm$0.02 &  0.81$\pm$0.02 \\
              LB~5959     & \texttt{659494049367276544} & 5.287  &0.043  &11.74$\pm$0.05  & $-$0.18$\pm$0.02 &  8.48$\pm$0.02 &  0.81$\pm$0.04 \\
              WD~0840+190 & \texttt{661010005319096192} & 5.135  &0.047  &11.81$\pm$0.03  & $-$0.13$\pm$0.02 &  8.42$\pm$0.02 &  0.76$\pm$0.02 \\
              WD~0833+198 & \texttt{662798086105290112} & 5.145  &0.039  &11.42$\pm$0.03  & $-$0.22$\pm$0.02 &  8.31$\pm$0.02 &  0.73$\pm$0.02 \\
              WD~0840+205 & \texttt{661841163095376896} & 5.376  &0.035  &11.90$\pm$0.03  & $-$0.11$\pm$0.02 &  8.61$\pm$0.03 &  0.78$\pm$0.02 \\
              WD~0837+218 & \texttt{665139697978259200} & 5.185  &0.039  &11.51$\pm$0.07  & $-$0.17$\pm$0.03 &  8.39$\pm$0.04 &  0.69$\pm$0.05 \\
              LB~8648     & \texttt{660178942032517760} & 5.370  &0.040  &11.81$\pm$0.03  & $-$0.13$\pm$0.02 &  8.54$\pm$0.02 &  0.77$\pm$0.03 \\
		\hline
	\end{tabular}
\end{table*}

These formal errors on both $M_{\rm f}$ and ${\rm log}(t_{\rm cool})$
are 2-3 times larger than for the Hyades WDs, because of larger error
bars on ${\rm M_G}$ and ${\rm (G_{BP}-G_{RP})_{0}}$, but still
comparable to the typical errors obtained when employing spectroscopic
measurements of $g-T_{\rm eff}$ pairs \citep[see, e.g.][for a recent
  analysis]{cummings18}.

\begin{figure}
	\includegraphics[width=\columnwidth]{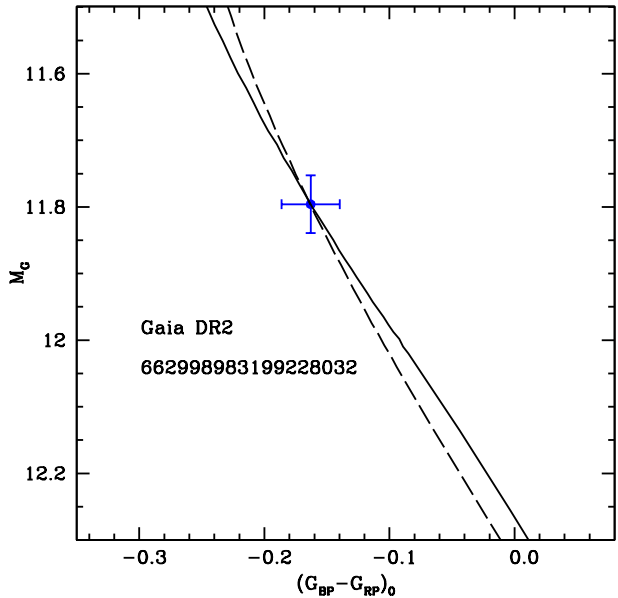}
        \caption{Best fit interpolated DA and DB cooling tracks for
          the new Praesepe WD \#662998983199228032 (see text for
          details). Error bars include the DR2 quoted photometric
          errors and the contribution from the parallax error.}
    \label{fig:new}
\end{figure}

As for the Hyades WDs of Paper~I, all Praesepe WDs in this sample have
evolved beyond the luminosity range where neutrino energy losses
dominate, but have not yet started crystallization.  Derivation of
$M_{\rm f}$ and $t_{\rm cool}$ employing the independent CO WD models
by \citet{fbb01} and \citet{renedo10}\footnote{Using the same
  bolometric corrections employed for the reference \citet{bastiwd}
  models. See Paper~I for a brief discussion on the main differences
  between these two sets of models and the \citet{bastiwd}
  calculations.}  provide masses unchanged compared to the values in
Table~\ref{tab:dataWD}. As for the case of the Hyades WDs, cooling
times obtained with \citet{fbb01} models are the same as the values of
Table~\ref{tab:dataWD}, whilst \citet{renedo10} models provide values
of $t_{\rm cool}$ within just dlog($t_{\rm cool}$)=$\pm$0.02 of the
results in Table~\ref{tab:dataWD}. These small differences have a
minor impact on the values of $M_{\rm i}$ derived in the next section.

Regarding the new WD \#662998983199228032, given that we do not have
spectroscopic information, we determined its mass and cooling age
$M_{\rm f}$ and $t_{\rm cool}$, by employing both H-atmosphere and
He-atmosphere ($M_{\rm He}/M_{\rm WD}=10^{-3.5}$) WD tracks from
\citet{bastiwd}. This object has a parallax (after zero point
correction) $\pi=5.201\pm 0.261$~mas, ${\rm M_G}=11.8\pm 0.04$
(including also the contribution of the parallax error), and colour
${\rm (G_{BP}-G_{RP})_{0}}=-0.16\pm 0.02$.  Figure~\ref{fig:new}
displays the best fit interpolated DA and DB tracks compared to the
CMD of the object. If this WD is of DA spectral type, we derive ${\rm
  log}(t_{\rm cool})=8.50\pm 0.04$ (age in years) and $M_{\rm
  f}=0.80\pm 0.04 M_{\odot}$. In case of DB spectral type we obtain
${\rm log}(t_{\rm cool})=8.60\pm 0.04$ and $M_{\rm f}=0.73\pm 0.05
M_{\odot}$. Given these non negligible uncertainties especially on the
cooling times, due to the unknown spectral type, we won't include this
object in the IFMR analysis that follows, although for the sake
  of comparison we will show its position in the $M_{\rm f}$ vs
  $M_{\rm i}$ diagram in Sect.~\ref{ifmr}, assuming it is of DA type
  like the other cluster WDs.

\begin{figure}
	\includegraphics[width=\columnwidth]{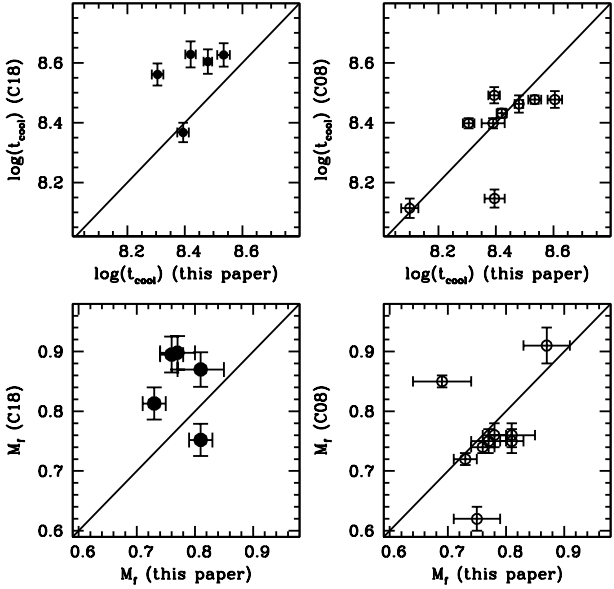}
        \caption{Comparison of $t_{\rm cool}$ (upper panels) and
          $M_{\rm f}$ (lower panels) between our results of
          Table~\ref{tab:dataWD} and the corresponding values from C18
          (left) and C08 (right), respectively. Solid lines display
          the 1:1 relationship between the quantities on the
          horizontal and vertical axis.}
    \label{fig:comp}
\end{figure}

Figure~\ref{fig:comp} compares our determination of WD masses and
cooling times with the results by \citet{catalan08} --hereafter C08--
and \citet{cummings18} --hereafter C18.  These two independent studies
employ $g$ and $T_{\rm eff}$ values for individual WDs from different
spectroscopic sources (see the papers for details), and make use of
the \citet{sala00} and \citet{fbb01} WD cooling tracks,
respectively. We have verified that also \citet{sala00} models agree
with the \citet{bastiwd} ones in terms of luminosity and $T_{\rm eff}$
time evolution for a given WD mass, in the relevant luminosity range.
Any difference in the $M_{\rm f}$ and $t_{\rm cool}$ compared to our
analysis is therefore due to inconsistencies between the $g$-$T_{\rm
  eff}$ pairs obtained from the WD spectroscopy, and the CMD location
of the WD sample.

All objects in our final WD sample (Table~\ref{tab:dataWD}) are also
in C08 study, whereas C18 include 5 objects common to our sample
(LB~1876 LB~5959 WD~0840+190 WD~0833+198 LB~8648).  On average C08
results are in agreement with ours, both in terms of $M_{\rm f}$ and
$t_{\rm cool}$. The most discrepant WD regarding $t_{\rm cool}$ is the
magnetic WD EGGR59, whose cooling time determined by C08 is about a
factor of 2 lower than our determination. Differences for the other
objects are smaller; for several objects cooling times are in
agreement with ours within the errors.

There is also only one object in C08 whose $M_{\rm f}$ value is
substantially higher than what we found, namely WD0837+218. C08
determined for this WD $M_{\rm f}$=0.85$\pm$0.01~$M_{\odot}$, whilst
we obtain 0.69$\pm$0.05~$M_{\odot}$.  On the other hand, four out of
five WDs in C18 sample display longer cooling times and higher masses
than in our analysis, namely LB~5959, WD~0840+190, WD~0833+198,
LB~8648.

\section{The initial-final mass relation}
\label{ifmr}

As discussed in the Introduction, 
having determined WD masses and cooling ages from the CMD, we
need a cluster age from the MS turn off to determine the initial values $M_{\rm i}$ for our WD sample.

\citet{gaiaclust} determined a turn off age log($t_{\rm
    cl}$)=8.85$^{+0.08}_{-0.06}$ ($t_{\rm cl}$ in years) obtained from
  \textit{Gaia} DR2 photometry and parallaxes, employing the PARSEC
  \citep{parsecmimf}\footnote{\citet{parsecmimf} isochrones are the
    PARSEC isochrones by \citet{parsec} extended to the end of the
    thermal pulse phase using the synthetic AGB technique (see the
    original paper for details).}  isochrones for [Fe/H]=0.14 --a
  metallicity consistent with spectroscopic measurements, see
  e.g. \citet{cdm17} and references therein-- transformed to the
  \textit{Gaia} DR2 photometric system.
Using this age (and error bar) and --consistently with the cluster age
estimate-- the initial mass-lifetime values from \citet{parsecmimf}
evolutionary tracks, we have determined $M_{\rm i}$ for our WD
sample, as listed in Table~\ref{tab:ifmr}. 

\begin{table}
	\centering
	\caption{Initial and final masses for the 9 DA WDs of
          Table~\ref{tab:dataWD}. From left to right we display the WD
          name, the initial mass (in solar masses), the asymmetric
          error bars estimated from the cooling times, and the final
          mass with associated error (in solar masses -- see
          Table~\ref{tab:dataWD}).}
	\label{tab:ifmr}
	\begin{tabular}{rcccc} 
		\hline
		Name & $M_{\rm i}$ & ${\Delta^{-}}$ & ${\Delta^{+}}$ &$M_{\rm f}\pm\sigma$ \\
		\hline
               LB~5893     & 2.78&    0.20 &  0.17 &0.87$\pm$0.04 \\
               EGGR~59     & 3.02&    0.28 &  0.26 &0.75$\pm$0.04 \\
               LB~1876     & 3.02&    0.28 &  0.25 &0.81$\pm$0.02 \\
               LB~5959     & 3.16&    0.32 &  0.31 &0.81$\pm$0.04 \\
               WD~0840+190 & 3.05&    0.29 &  0.27 &0.76$\pm$0.02 \\
               WD~0833+198 & 2.92&    0.24 &  0.22 &0.73$\pm$0.02 \\
               WD~0840+205 & 3.50&    0.46 &  0.51 &0.78$\pm$0.02 \\
               WD~0837+218 & 3.01&    0.28 &  0.26 &0.69$\pm$0.05 \\
               LB~8648     & 3.28&    0.37 &  0.37 &0.77$\pm$0.03 \\
		\hline
	\end{tabular}
\end{table}

\begin{figure}
	\includegraphics[width=\columnwidth]{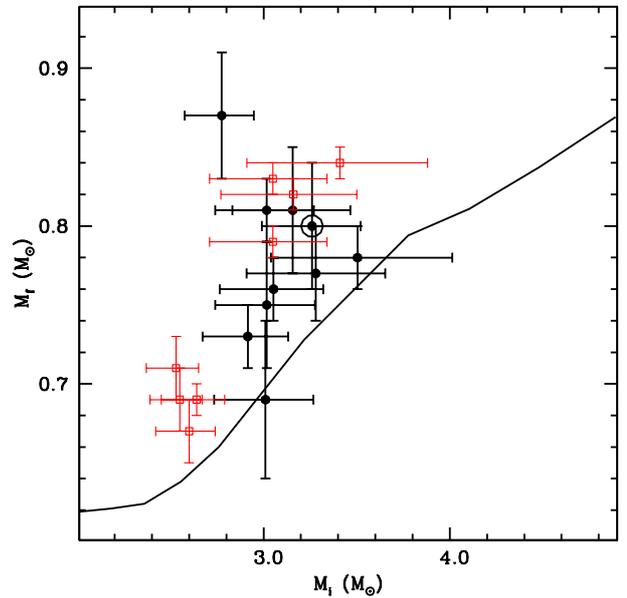}
        \caption{IFMR for Praesepe (filled circles with error bars,
          from Table~\ref{tab:ifmr}).  For the sake of comparison we
          display also the IFMR for the Hyades derived in Paper~I
          (open squares with error bars). The solid line displays the
          IFMR from the \citet{parsecmimf} theoretical models. 
            The open circle highlights the new cluster WD
            \#662998983199228032, in case it is of DA spectral type,
            like the others.}
    \label{fig:ifmr}
\end{figure}

Figure~\ref{fig:ifmr} shows the IFMRs from the data in
Table~\ref{tab:ifmr}, plus the Hyades IFMR from Paper~I. As well
known, the Hyades cluster has basically the same [Fe/H] of Praesepe
\citep[see][and references therein]{cdm17} and a very similar age
\citep[in Paper~I we used the the DR2 determinations for the Hyades
  age, log($t_{\rm cl}$)=8.90$^{+0.08}_{-0.06}$, see][]{gaiaclust}.

Errors in $M_{\rm i}$ for Praesepe range between $\sim$0.2 and
$\sim$0.5~$M_{\odot}$, with typical values equal to
0.2-0.3~$M_{\odot}$.  Like for the Hyades WDs studied in Paper~I,
errors on $M_{\rm i}$ are dominated by the error bar on the cluster
age, hence this is essentially a systematic error on all WD initial
masses, because increasing or decreasing the cluster age according to
its error bar does systematically decrease or increase, respectively,
the values of $M_{\rm i}$ for all WDs of any $M_{\rm f}$.

As we discuss below, Hyades and Praesepe IFMRs are in good agreement,
but is this agreement preserved when considering the systematic error
bars on $M_{\rm i}$? For example, if we adopt the lower values of the
Hyades WD initial masses and the upper values of Praesepe ones --that
means, the upper limit of the Hyades age determination, and the lower
limit of Praesepe age according to \citet{gaiaclust}-- there is
clearly a systematic shift between the two IFMRs.  However, the errors
on the Hyades and Praesepe ages, that are determined by the
uncertainty in fitting the sparsely populated turn offs
\citep{gaiaclust}, cannot be considered independently.

\begin{figure*}
\begin{center}
        \includegraphics[width=135mm]{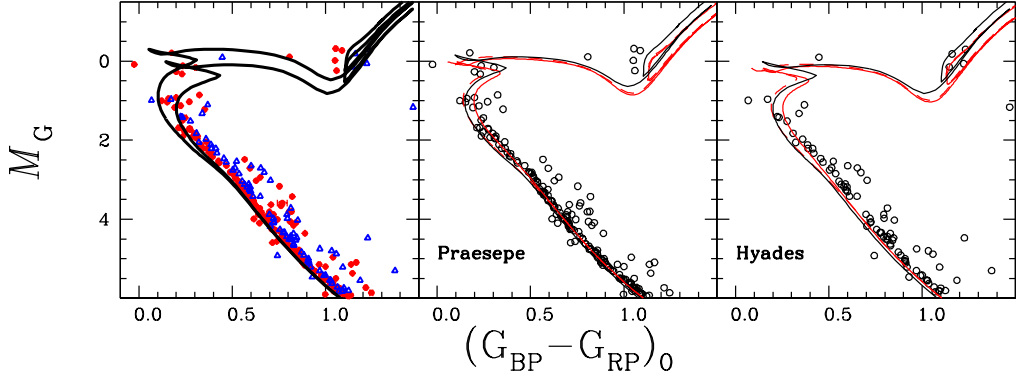}
        \caption{\textit{Gaia} DR2 CMDs of the Hyades (blue empty
          triangles) and Praesepe (red filled circles), together with
          -- black solid lines-- the best fit \citet{parsecmimf}
          isochrones for log(t)=8.85 and log(t)=8.90 (left panel --
          see text for details). In the middle and right panels we
          display separately Praesepe and Hyades CMDs together with
          the best fit PARSEC isochrones (black solid lines) .  Red
          dashed and solid lines in the middle and bottom panels
          display non-rotating and rotating MIST isochrones,
          respectively.  In case of Praesepe MIST isochrones have ages
          equal to log(t)=8.80 (non rotating) and log(t)=8.82
          (rotating), whilst for the Hyades MIST isochrones have ages
          equal to log(t)=8.85 (non rotating) and log(t)=8.87
          (rotating -- see text for details).}
    \label{fig:age}
\end{center}
\end{figure*}

For example, the upper age limit for Praesepe is larger than the lower
age limit for the Hyades, but Praesepe cannot be older than the
Hyades, when looking at the top panel of Fig.~\ref{fig:age}, where the
CMDs (distance and reddening corrected) of the two clusters are
superimposed, together with the best fit isochrones from
\citet{gaiaclust}.  Praesepe turn-off absolute magnitude is clearly
not fainter than the Hyades one.

Instead, when considering the errors on the cluster ages, the
difference between the best-fit ages of Hyades and Praesepe should be
preserved. As a consequence, if we consider for example the lower
limits of the $M_{\rm i}$ values for the Hyades given by their error
bars --that is, if we consider the upper limit of the Hyades age
determination-- we should at the same time consider also the lower
limit of Praesepe initial masses.

Bearing in mind this discussion on the cluster ages' errors,
Fig.~\ref{fig:ifmr} shows that the IFMRs of these two clusters appear
broadly consistent, even when considering the error bars on $M_{\rm
  i}$.  The only discrepant object is Praesepe WD LB~5893, with
$M_{\rm i}=2.78^{+0.17}_{-0.20}$ for a WD mass $M_{\rm f}=0.87\pm
0.04$ -- see also Fig.~\ref{fig:sample} for its \textit{anomalous}
position along the cooling sequence.  Even if we neglect LB~5893, and
as noticed in Paper~I for the Hyades WDs alone, there is a hint of a
small spread in the values of $M_{\rm f}$, for initial masses around
3~$M_{\odot}$.  Notice that the IFMR of the new WD
  \#662998983199228032 (that we do not consider in the analysis that
  follows) is consistent with the bulk of the other WDs, if it is
  assumed to be of spectral type DA like the others.
 
To quantify better the general agreement between the two cluster
IFMRs, we considered $M_{\rm i}$ and $M_{\rm f}$ values for Praesepe
WDs, the error bars on $M_{\rm f}$ reported in Table~\ref{tab:ifmr},
and the random errors on $M_{\rm i}$ due to the error on the WD
cooling times only. For the Hyades we considered the same quantities,
as obtained from the data in Paper~I.  Typical random errors on
$M_{\rm i}$ are 0.01$M_{\odot}$ for the Hyades, and just a few
hundredths of solar masses for Praesepe WDs.  A linear fit to the
$M_{\rm i}$-$M_{\rm f}$ data for both clusters (17~WDs), and
considering errors on both axis\footnote{We used the routine
  \textit{fitexy} from \citet{numrec}}, provides a slope $\Delta
M_{\rm f}/ \Delta M_{\rm i}$=0.20$\pm$0.02.  Discarding the Praesepe
WD LB~5893 does not change the result within the errors.  If we then
consider only the Hyades sample of 8 objects, we obtain $\Delta M_{\rm
  f}/ \Delta M_{\rm i}$=0.21$\pm$0.02,  consistent with the full
sample.  Also the zero points are consistent within the errors when
considering the Hyades sample (0.16$\pm$0.05 $M_{\odot}$) and the
combined Hyades+Praesepe one (0.20$\pm$0.05 $M_{\odot}$).  This
agreement of the IFMR for the two clusters is preserved also when
considering the lower or the upper limits on $M_{\rm i}$ for the two
clusters (see previous discussion about the error bar on the cluster
ages).

The theoretical IFMR by \citet{parsecmimf} is also displayed in
Fig.~\ref{fig:ifmr}. The slope within the lower and upper $M_{\rm i}$
limits for Hyades and Praesepe WDs is slightly shallower than the data
($\Delta M_{\rm f}/ \Delta M_{\rm i} \sim$0.13), and on the whole this
theoretical IFMR underestimates $M_{\rm f}$ for the initial mass range
covered by these two clusters, even when considering the systematic
error bars on $M_{\rm i}$.
 
We have also tested the self-consistent use of an independent set of
stellar evolution calculations.  To this purpose we have downloaded
MIST stellar evolution tracks and isochrones \citep{mist} for
[Fe/H]=0.14, from the web-interpolator of the MIST stellar evolution
database\footnote{\url{http://waps.cfa.harvard.edu/MIST/}}, with
(initial linear and angular velocities equal to 40\% of the critical
values) and without rotation, and compared these isochrones with the
best-fit \citet{parsecmimf} isochrones employed by \citet{gaiaclust}
to determine Hyades and Praesepe ages.  This analysis complements
  and expands upon the work by C18, who compared IFMRs derived using
  PARSEC an MIST non-rotating isochrones for their star cluster
  sample, including Hyades and Praesepe.

Figure~\ref{fig:age} displays this comparison, together with the
\textit{Gaia} DR2 CMDs of the Hyades and Praesepe. We show the MIST
isochrones that match the MS turn-off luminosity of the
\citet{parsecmimf} counterpart.  We remark that it is not the purpose
of this paper to discuss the age-determination for these two clusters,
but just to select the age of MIST isochrones that match the turn-off
absolute G-magnitude of the best-fit isochrones determined by
\citet{gaiaclust}.

We find that the cluster ages obtained by \citet{gaiaclust} would be
0.05~dex (about 80~Myr) and 0.03~dex (about 50~Myr) younger when using
the non-rotating and rotating MIST isochrones,
respectively\footnote{Error bars on the MIST cluster ages are the same
  as in \citet{gaiaclust}, because age differences between isochrones
  with the same turn-off absolute magnitudes are constant within the
  age ranges spanned by \citet{gaiaclust} error bars.}. With these
ages we then employed the MIST initial mass-lifetime values from
\citet{mist} calculations, and the cooling times derived above, to
determine $M_{\rm i}$ values and the IFMR of Hyades and Praesepe.

\begin{figure}
	\includegraphics[width=\columnwidth]{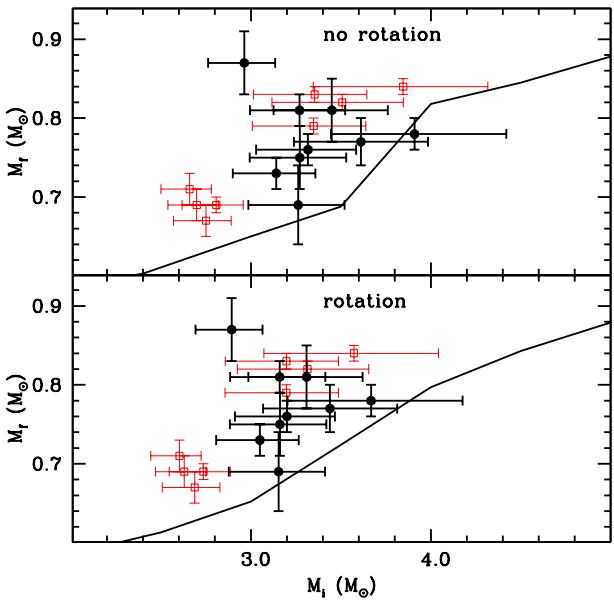}
        \caption{As Fig.~\ref{fig:ifmr}, but considering cluster ages
          and progenitor lifetimes from the \citet{mist} models with
          (lower panel) and without (upper panel) the inclusion of
          rotation. Solid lines denote the corresponding theoretical
          IFMRs.}
    \label{fig:ifmr_b}
\end{figure}

Compared to the reference IFMR of Fig.~\ref{fig:ifmr}, the MIST based
IFMRs display $M_{\rm i}$ values typically larger at a given WD mass,
with differences varying from object to object and also from cluster
to cluster. On average, initial masses $M_{\rm i}$ are larger by
$\sim$0.10-0.15~$M_{\odot}$ for both Hyades and Praesepe when
considering the rotating MIST models, while these differences increase
to on average $\sim$0.2-0.3~$M_{\odot}$ with the non-rotating models.
Figure~\ref{fig:ifmr_b} shows the resulting IFMRs, compared to the
theoretical counterpart obtained from the same calculations. Also in
case of using self-consistently MIST models, the theoretical IFMR on
average underpredicts the WD masses, even considering the systematic
errors on $M_{\rm i}$ due to the relatively large errors on the
cluster ages.  The slopes of the semi-empirical IFMR --calculated as
described before-- are slightly changed compared to the case of the
reference IFMR only for the case of non-rotating models.  When
considering just the Hyades sample, rotating MIST models give $\Delta
M_{\rm f}/ \Delta M_{\rm i}$=0.19$\pm$0.01, whilst this slope becomes
$\Delta M_{\rm f}/ \Delta M_{\rm i}$=0.18$\pm$0.01 for the combined
Hyades+Praesepe sample.  Non-rotating MIST models give instead $\Delta
M_{\rm f}/ \Delta M_{\rm i}$=0.15$\pm$0.01 for the Hyades and $\Delta
M_{\rm f}/ \Delta M_{\rm i}$=0.14$\pm$0.01 for the two clusters
combined.  These are slightly shallower slopes than the reference
IFMR.  As for the reference IFMR, the zero points are consistent
between the Hyades only and the Hyades+Praesepe samples, for both
rotating and non-rotating MIST models.
 
\subsection{PRAESEPE AND HYADES IFMR EMPLOYING C18 CLUSTER PARAMETERS}
\label{c18}

The recent analysis by C18 has derived Hyades and Praesepe ages
  --obtained by $BV$ CMD fitting employing both PARSEC and
  non-rotating MIST isochrones-- with a much smaller formal error
  compared to \citet{gaiaclust} results. C18 distance moduli and
  [Fe/H] for both clusters are consistent with \citet{gaiaclust}, the
  only difference being $E(B-V)$=0 for Praesepe\footnote{Our Paper~I
    and C18 employ $E(B-V)$=0.0 for the Hyades. A zero reddening for
    the Hyades is confirmed also by the recent \citet{taylor}
    analysis.}  -- the \textit{traditional} value for Praesepe--
  instead of $E(B-V)$=0.027 \citep[see also][]{cdm17}. Hyades and
  Praesepe ages determined with the PARSEC models by C18 are equal to
  700$\pm$25~Myr and 705$\pm$25~Myr, respectively. When using
  non-rotating MIST models C18 give 705$\pm$50~Myr for the Hyades and
  685$\pm$25~Myr for Praesepe.  Notice that C18 ages are typically
  younger than our adopted values for the Hyades, whereas they are the
  same or older for Praesepe, but always within the large error bars
  of the ages based on the \textit{Gaia} DR2 parallaxes and CMDs.
   
Following the referee suggestion we have rederived the IFMR for both
Hyades and Prasepe using C18 results, \textit{Gaia}\,DR2 WD parallaxes
and photometry, employing both PARSEC and non-rotating MIST models to
determine progenitor ages and masses. This will enable us to make a
fully consistent comparison with C18 IFMR results, that made use of
spectroscopic estimates of WD masses and cooling times.

\begin{figure}
	\includegraphics[width=\columnwidth]{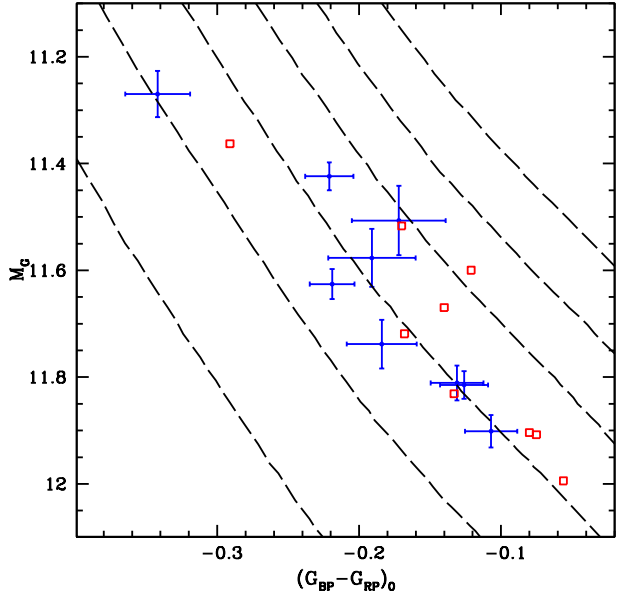}
        \caption{As Fig.~\ref{fig:sample}, but showing just the 9
          Praesepe WDs employed to determine the IFMR, together with
          the adopted cooling tracks.  Dots with error bars display
          the WD CMD when employing the reference reddening
          $E(B-V)$=0.027, while open squares (without error bars, for
          the sake of clarity) display the CMD assuming zero
          reddening, following C18 analysis (see text for details)}
    \label{fig:samplecumm}
\end{figure}

\begin{figure}
	\includegraphics[width=\columnwidth]{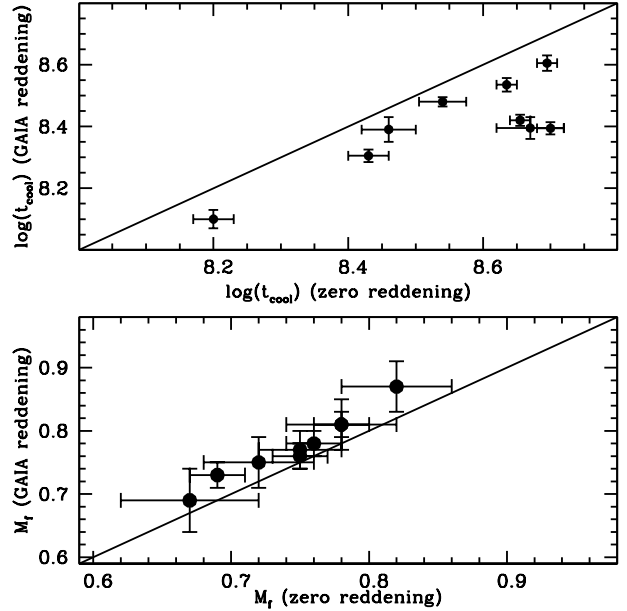}
        \caption{As Fig.~\ref{fig:comp}, but comparing the results in
          Table~\ref{tab:dataWD} with WD masses and cooling times
          obtained assuming zero reddening for Praesepe.}
    \label{fig:compcumm}
\end{figure}

Hyades WD masses and cooling times from \textit{Gaia}\,DR2 are
unchanged compared to the results of Paper~I, but this is not the case
for Praesepe.  A zero reddening instead of $E(B-V)$=0.027 implies
redder and fainter WDs compared to the results in
Table~\ref{tab:dataWD}.  To give an idea of the magnitude of this
effect, the reddening law by \citet{gaiaclust} gives, in the color
range of Praesepe WDs, ${\rm A_{\rm G}}\sim 3.1~E(B-V)$, and $E({\rm
  G_{BP}-G{RP}})\sim 1.7 E(B-V)$.  Figure~\ref{fig:samplecumm}
compares the distance and reddening corrected CMD of the 9 Praesepe
WDs employed in our IFMR determination, with the CMD of the same
objects but in case of $E(B-V)$=0.  Assuming zero reddening for
Praesepe as in C18 causes a systematic decrease of the derived WD
masses (by 0.02-0.05$M_{\odot}$), and a systematic increase of their
cooling ages (by 0.06-0.30~dex) compared to the values reported in
Table~\ref{tab:dataWD}, as shown in Fig.~\ref{fig:compcumm}

\begin{figure}
	\includegraphics[width=\columnwidth]{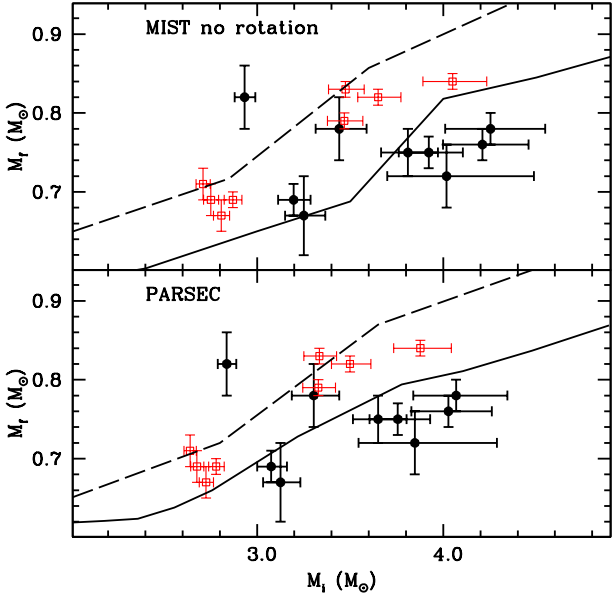}
        \caption{As Fig.~\ref{fig:ifmr_b}, but showing our IFMRs
          derived using ages and reddenings from C18. The dashed lines
          display C18 analytical IFMRs determined using PARSEC models
          (lower panel) and non-rotating MIST models (upper panel),
          whilst solid lines show the theoretical IFMRs predicted by
          PARSEC and MIST models (see text for details).}
    \label{fig:ifmrcumm}
\end{figure}

By employing C18 cluster ages we have then determined the Hyades and
Praesepe IFMRs, with both PARSEC and non-rotating MIST models.
Table~\ref{tab:ifmrcumm} summarizes the results, that are also
displayed in Fig.~\ref{fig:ifmrcumm}.

Both cluster IFMRs have now smaller errors associated to $M_{\rm i}$,
reflecting the much reduced error on the cluster ages.  The Hyades
IFMR is similar to our results in Figs.~\ref{fig:ifmr} and
~\ref{fig:ifmr_b} (see also Paper~I), with just a small systematic
increase of $M_{\rm i}$ by 0.1-0.4~$M_{\rm i}$, due to the lower
adopted cluster age. The result is however within the error bars of
Figs.~\ref{fig:ifmr} and ~\ref{fig:ifmr_b}.

The situation is quite different for Praesepe. Both $M_{\rm i}$ and,
to a smaller degree, $M_{\rm f}$ values are reduced compared to the
results in Figs.~\ref{fig:ifmr} and ~\ref{fig:ifmr_b}. The large
reduction of $M_{\rm i}$ values is mainly due to the sizably longer WD
cooling times compared to the results in Table~\ref{tab:dataWD}. There
is now an average offset between Praesepe and Hyades IFMR, with
Praesepe IFMR on average shifted towards larger values of $M_{\rm i}$
at fixed $M_{\rm f}$. This is inconsistent with C18 results and our
IFMRs determined employing \citet{gaiaclust} cluster parameters.
 
This difference between Hyades and Praesepe IFMRs is a consequence of
the assumption of zero reddening for Praesepe, and is not due to the
cluster ages determined by C18. If we employ just C18 cluster ages but
Praesepe WD masses and cooling times of Table~\ref{tab:dataWD}, we
still have consistency between Hyades and Praesepe IFMRs.

\begin{table}
	\centering
	\caption{As Table~\ref{tab:ifmr} but for Praesepe and Hyades
          IFMRs obtained adopting cluster ages and reddenings from C18
          (see text for details).}
	\label{tab:ifmrcumm}
	\begin{tabular}{rcccc} 
		\hline
		Name & $M_{\rm i}$ & ${\Delta^{-}}$ & ${\Delta^{+}}$ &$M_{\rm f}\pm\sigma$ \\
		\hline
                Praesepe & & & & \\
                \hline
               LB~5893     & 2.84&    0.05 &  0.05 &0.82$\pm$0.04 \\
               EGGR~59     & 3.85&    0.30 &  0.44 &0.72$\pm$0.04 \\
               LB~1876     & 4.07&    0.23 &  0.28 &0.78$\pm$0.02 \\
               LB~5959     & 3.30&    0.12 &  0.14 &0.78$\pm$0.04 \\
               WD~0840+190 & 3.76&    0.15 &  0.17 &0.75$\pm$0.02 \\
               WD~0833+198 & 3.08&    0.08 &  0.08 &0.69$\pm$0.02 \\
               WD~0840+205 & 4.03&    0.20 &  0.23 &0.76$\pm$0.02 \\
               WD~0837+218 & 3.13&    0.09 &  0.11 &0.67$\pm$0.05 \\
               LB~8648     & 3.65&    0.14 &  0.15 &0.75$\pm$0.03 \\
		\hline
                Hyades & & & & \\
                \hline
              HZ~14   & 2.64 & 0.03 &  0.03 & 0.71$\pm$0.02\\
              LAWD~19 & 2.68 & 0.03 &  0.04 & 0.69$\pm$0.02\\
              HZ~7    & 2.73 & 0.04 &  0.04 & 0.67$\pm$0.02\\
              LAWD~18 & 2.78 & 0.04 &  0.04 & 0.69$\pm$0.01\\
              HZ~4    & 3.33 & 0.08 &  0.09 & 0.79$\pm$0.01\\
              EGGR~29 & 3.34 & 0.08 &  0.09 & 0.83$\pm$0.01\\
              HG~7-85 & 3.50 & 0.10 &  0.11 & 0.82$\pm$0.01\\
              GD~52   & 3.88 & 0.14 &  0.17 & 0.84$\pm$0.01\\
		\hline
	\end{tabular}
\end{table}

Figure~\ref{fig:ifmrcumm} displays also the analytical IFMRs by C18
obtained with both PARSEC and non-rotating MIST models. These C18
IFMRs show only a small offset (by $\sim$0.1~$M_{\odot}$) towards
smaller $M_{\rm i}$ values at fixed $M_{\rm f}$ when compared to our
Hyades results. This is mainly due to the slightly longer cooling
times we obtain for the Hyades WDs, compared to C18.  Praesepe results
are instead completely inconsistent with C18 IFMRs.

When compared to the theoretical counterparts from PARSEC and
non-rotating MIST models --also displayed in Fig.~\ref{fig:ifmrcumm}--
our semi-empirical Hyades results predict larger $M_{\rm f}$ values
for the initial mass range covered by the cluster. This is in
agreement with C18 and our IFMRs in Figs.~\ref{fig:ifmr} and
\ref{fig:ifmr_b}.  In case of Praesepe most of the WDs now lie below
the theoretical IFMRs in Fig.~\ref{fig:ifmrcumm}, implying lower
$M_{\rm f}$ values than predicted from theory.

The reddening adopted for Praesepe is therefore crucial to establish
photometrically the cluster IFMR.  The value of $E(B-V)$=0.027
determined by \citet{gaiaclust} from isochrone fitting is in agreement
with independent estimates by \citet{taylor}. When employing this
reddening and the extinction law from \citet{gaiaclust}, the Praesepe
IFMR is consistent with the Hyades result.  Assuming the traditional
zero reddening for the \textit{Gaia} CMD of Praesepe WDs, induces a
very large dispersion in the global IFMR obtained from these two
clusters that are almost coeval and with the same metallicity.  
\section{Summary and discussion}
\label{conclusions}

We have employed the \textit{Gaia}\,DR2 sample of \textit{bona-fide}
Praesepe member stars, and selected the objects on the WD cooling
sequence.  Ten out of a total of 12 WDs --all of spectral type DA--
satisfy the quality criteria selected from the quality indicators
available in \textit{Gaia}\,DR2.  Nine objects are already known DA
WDs, the remaining one is a new WD member, with DR2 identifier
\#662998983199228032. A spectroscopic follow-up is needed to establish
the spectral type of this new object.

We have determined masses and cooling times of the WDs by matching
their CMD \citep[corrected for reddening and distance using
    reddening and parallaxes from][]{gaiaclust} with theoretical
cooling sequences.  The accuracy of DR2 parallaxes ($\sim$4\%
fractional errors) and photometry (errors of a few hundredths of a
magnitude) has allowed us to determine masses with an average error of
3--5\%, and cooling times with an average uncertainty of 5--7\%.  For
the new WD of unknown spectral type, we derive two pairs of cooling
time-mass values, namely ${\rm log}(t_{\rm cool})=8.50\pm 0.04$,
$M_{\rm f}=0.80\pm 0.04 M_{\odot}$ from DA tracks, and ${\rm
  log}(t_{\rm cool})=8.60\pm 0.04$ and $M_{\rm f}=0.73\pm 0.05
M_{\odot}$ from DB tracks.

An IFMR for the confirmed DA Praesepe WDs in our sample has been then
determined by assuming the cluster MS turn-off age ($\sim$710~Myr)
recently determined from the \textit{Gaia} DR2 cluster CMD and PARSEC
stellar evolution models \citep{gaiaclust}.  This Praesepe IFMR is
consistent with the Hyades IFMR derived in Paper~I from DR2 data and
the same methods applied here.  We have also derived self-consistently
Praesepe and Hyades IFMRs employing two alternative sets of stellar
evolution models, the non-rotating and rotating MIST models.  The use
of MIST models shifts the individual $M_{\rm i}$ values towards higher
values compared to the reference PARSEC results, due to a different
progenitor mass-lifetime relationship, and the typically younger ages
derived with MIST models. The magnitude of these shifts depend on
whether the non-rotating (increase by 0.20-0.30~$M_{\odot}$) or
rotating (increase 0.10-0.15~$M_{\odot}$) MIST models are employed.

In these IFMR determinations, the Praesepe WD LB~5893 appears to
deviate from the rest of the objects in the $M_{\rm i}-M_{\rm f}$
diagram, resulting too massive for its derived $M_{\rm i}$. This is
consistent with its \textit{anomalous} location along the cooling
sequence (see Fig.~\ref{fig:sample}), and confirms previous findings
\citep[see][]{clav01, casewell} based on spectroscopic $g-T_{\rm eff}$
determinations.  As discussed in \citet{casewell}, there is nothing
peculiar about this object, for neither magnetic fields nor rapid
rotation was detected.  These authors speculate whether it may have
formed from a blue straggler star, given the known presence of blue
stragglers in this cluster \citep[e.g., ][and references
  therein]{fossati}. Obviously strong differential mass loss is also a
possibility, even though among the known cluster WDs we see such a
strong effect only for this single object.  We also notice that
  this star is the most massive WD detected so far in Praesepe, and is
  located near the cluster centre. It is possible that close
  interactions between massive stars in the denser cluster core might
  have affected the IFMR for this WD.

Two other interesting objects are LB~5959 and EGGR~59. Regarding
  LB~5959, we have already mentioned in Sect.~\ref{data} that
  \citet{casewell12} observations suggest the presence of a companion
  with mass equal to 25-30~$M_{\rm Jup}$, although \textit{Gaia}\,DR2
  parameters are consistent with this object being a single star. In
  the scenario envisaged by \citet{casewell12} the substellar
  companion must have been engulfed by the WD progenitor during the
  AGB evolution. This common envelope interaction may therefore have
  modified the IFMR of this object, compared to our estimates based on
  single-star evolution for the progenitor.
 
As for the magnetic WD EGGR~59, the origin of WDs with strong
magnetic fields is still debated \citep[see, e.g.][for
  reviews]{ferr15, magnetic16}.  These fields might be fossil, the
remnants of original weak magnetic fields amplified during the course
of the evolution of the progenitor \citep[][]{abl81}.  According to a
competing scenario \citep{tout}, all highly magnetic white dwarfs
(defined as WDs with fields in excess of 1~MG, like EGGR~59), both
single stars or the components of magnetic cataclysmic variables, have
instead a binary origin. Interestingly, also EGGR~59 is located
  in the cluster central region, where interactions between stars are
  more likely.

As shown in Fig.~\ref{fig:sample}, this star sits nicely within the
Praesepe cooling sequence; assuming that magnetic fields do not affect
the WD mass-radius relation and the bolometric corrections for the
\textit{Gaia} photometric filters, its mass is fully consistent with
the general IFMR of the other non-magnetic WDs.  Regarding the WD
mass-radius relation, a fundamental problem is that the surface
magnetic field of a star does not necessarily reflect the internal
field. According to the models by \citet{sm00}, very high internal
magnetic fields, of the order of $10^{11}-10^{13}~G$, can modify the
non-magnetic mass-radius relation, resulting in increased radii at
fixed WD mass.  This would cause an underestimate of the WD mass from
CMD analyses, when non-magnetic WD models are employed.  On the other
hand, assuming the mass-radius relation is unaffected by the internal
(unknown) magnetic field strength, a recent study by \citet{kulebi},
who applied magnetized WD spectra to infer mass and cooling times of
this object, found that within the errors of their diagnostics this WD
does not significantly deviate from the mean IFMR of non-magnetized
WDs.  If this result is confirmed by more comprehensive analyses of
the effect of magnetic fields on WDs (both evolutionary and spectral
properties), it could help constraining the scenario for magnetic WD
formation.

The comparison of our Hyades+Praesepe IFMR with theoretical
predictions also discloses -- confirming an analogous result by C18
(see their Fig.~5)-- a systematic discrepancy between theoretical IFMR
predictions and the semi-empirical results. Both PARSEC and MIST
(rotating and non-rotating) calculations --that include the full AGB
evolution-- on average do underpredict the final WD masses in the
initial mass range covered by these clusters. The size of the
discrepancy is of the order of a few 0.01~$M_{\odot}$, the larger
discrepancy found for the IFMR determined with the MIST rotating
models. This sets important constraints on the growth of the CO core
in AGB stars with these progenitor masses, with implications for the
efficiency of mass loss, third dredge-up, and contribution of AGB
stars to the integrated infrared light of stellar populations.

The photometric determination of Praesepe IFMR --and more in
  general for all \textit{Gaia} clusters with precise parallax and
  magnitude measurements of their WD populations-- relies crucially on
  the reddening assumed for the cluster.  All these results are based
  on employing for consistency a value $E(B-V)$=0.027 determined by
  \citet{gaiaclust} from isochrone fitting to the \textit{Gaia}\,DR2
  CMD of main sequence, turn off and core He-burning stars. This value
  is in agreement with $E(B-V)$=0.027$\pm$0.004 obtained independently
  by \citet{taylor}.

We have also determined Praesepe WD masses and cooling times employing
the traditional zero reddening for Praesepe, as in C18 analysis. In
this case WD masses are reduced and cooling times largely increased
compared to the case of $E(B-V)$=0.027.  This leads to a very
different IFMR \citep[irrespectively of using][or C18 cluster age
  estimates]{gaiaclust}, that is on average shifted to larger $M_{\rm
  i}$ values compared to the Hyades, implying a large dispersion in
the IFMR even for two clusters with the same metallicity and
approximately the same age.

\section*{Acknowledgments}
We are deeply indebted to Pierre Bergeron who kindly provided us with
bolometric corrections to the \textit{Gaia}~DR2 system for the WD
cooling tracks.  We thank our referee for comments that have
  improved our analysis and presentation of the results.  This work
presents results from the European Space Agency (ESA) space mission
\textit{Gaia}.  Gaia data are being processed by the \textit{Gaia}
Data Processing and Analysis Consortium (DPAC).  Funding for the DPAC
is provided by national institutions, in particular the institutions
participating in the \textit{Gaia} MultiLateral Agreement (MLA). The
\textit{Gaia} mission website is https://www.cosmos.esa.int/gaia.  The
\textit{Gaia} archive website is https://archives.esac.esa.int/gaia.



\bibliographystyle{mnras}
\bibliography{Praesepe_MiMf} 








\bsp	
\label{lastpage}
\end{document}